\documentclass[12pt]{article}
\usepackage{geometry}
\usepackage{a4}
\usepackage{graphicx}
\usepackage{epsf}
\usepackage{amsmath}
\usepackage{amssymb}
\usepackage{cite}
\newcommand{\be}{\begin{equation}}
\newcommand{\ee}{\end{equation}}

\newcommand{\Rmnum}[1]{\expandafter\@slowromancap\romannumeral #1@}
\newcommand{\bea}{\begin{eqnarray}}
\newcommand{\eea}{\end{eqnarray}}
\begin{document}
\def\C{{\mathbb{C}}}
\def\R{{\mathbb{R}}}
\def\s{{\mathbb{S}}}
\def\T{{\mathbb{T}}}
\def\Z{{\mathbb{Z}}}
\def\W{{\mathbb{W}}}
\def\Bbb{\mathbb}
\def\BZ{\Bbb Z} \def\BR{\Bbb R}
\def\BW{\Bbb W}
\def\BM{\Bbb M}
\def\BC{\Bbb C} \def\BP{\Bbb P}
\def\CP{\BC\BP}
\begin{titlepage}
\title{Geodesic Congruences and Their Deformations in Bertrand
  Space-times} \author{} 
\date{%
Prashant Kumar, Kaushik Bhattacharya, Tapobrata Sarkar 
\thanks{\noindent 
E-mail:~ kprash, kaushik, tapo@iitk.ac.in} 
\vskip0.4cm 
{\sl Department of Physics, \\ 
Indian Institute of Technology,\\ 
Kanpur 208016, \\ 
India}} 
\maketitle 
\abstract{We study the energy conditions and geodesic deformations in
Bertrand space-times. We show that these 
can be thought of as interesting physical space-times in
certain regions of the underlying parameter space, where the weak and strong
energy conditions hold. We further compute the ESR parameters and
analyze them numerically. The focusing of radial time-like and radial
null geodesics is shown explicitly, which verifies the Raychaudhuri equation.
\noindent
}
\end{titlepage}
\section{Introduction}
The Schwarzschild metric, discovered nearly a century ago, remains
one of the simplest yet most profound solutions of Einstein's field
equations. One of the reasons for the popularity of the Schwarzschild
solution among relativists is that it provides a realistic scenario to
describe closed, stable orbits of planets and other heavenly
objects. It is however well known that there are other solutions of
Einstein's equations which can also describe such stable, periodic
motion. One class of examples was discovered in a remarkable paper by
Perlick \cite{perlick}, and these were named ``Bertrand space-times''
(BSTs). Indeed, Perlick's classification generalizes the well known
Bertrand's theorem \cite{bert} in Newtonian mechanics (an excellent exposition can be found in \cite{goldstein}) 
to general relativity.  The
former theorem states that the
harmonic oscillator and Kepler potentials are the only
spherically symmetric potentials for which bounded orbits are
periodic. Based on the standard deductions of the Bertrand's theorem 
there was an attempt to generalize its form in the special relativistic case 
\cite{Kumar}. Perlick's work determines all static, spherically symmetric
space-times in the most general case where one can have stable, closed orbits.

Apart from being interesting from a purely theoretical perspective,
BSTs might also be relevant for other reasons. For example, one
possibility may be to model a realistic space-time that allows for
bounded, periodic orbits. Although such a possibility was ruled out in
Perlick's original work due to the fact that asymptotically flat BSTs
(relevant for the motion of objects around an isolated mass) do not
seem to satisfy the weak energy condition (WEC) at infinity,
asymptotically non-flat BSTs are equally interesting objects, as
alternatives to black hole space-times.

This paper studies a class of BSTs from the perspective of the energy
conditions and geodesic deformations. We analyze these aspects and
find that in certain regions of the parameter space, BSTs do obey the
strong and weak energy conditions. We further study the Raychaudhuri
equation for BSTs and confirm the geodesic focusing theorem.

The paper is organized as follows. In the next section, we briefly
review BSTs and analyze the energy conditions
therein. In section 3, we study radial and circular geodesic flows in
a class of BSTs and analyze the focusing theorem. Section 4 ends with
our conclusions and directions for further study.
\section{Energy Conditions in Bertrand space-times}
\label{bst}
Formally, the definition of a Bertrand space-time 
\cite{perlick}, \cite{enciso} arises via a static, spherically
symmetric Lorentzian manifold $(M, g)$ whose domain is diffeomorphic
to a product manifold $(r_1\,,\,r_2)\times S^2\times \mathbb{R}$ with
the metric $g$ given by
\begin{eqnarray}
ds^2 = -e^{2\nu(r)} dt^2 + e^{2\lambda(r)} dr^2 + r^2 (d \theta^2 +
\sin^2 \theta d \phi^2)\,,
\label{invl}
\end{eqnarray}
where $r$ ranges in the open interval $(r_1\,,\,r_2)$, $\theta$ and
$\phi$ are co-ordinates on the two-sphere.  $\lambda$ and $\nu$ are some unspecified functions 
of $r$ to start with. Such a Lorentzian manifold is called a BST provided there is a
circular trajectory passing through each point in the interval
$(r_1\,,\,r_2)$, which is stable under small perturbations of the
initial conditions.

Starting from this definition, Perlick \cite{perlick} deduced that there can be two categories of BSTs given by:
\begin{eqnarray}
ds^2 &=&-\frac{dt^2}{G\mp r^2[1- Dr^2\pm \sqrt{(1-Dr^2)^2 - K
r^4}]^{-1}}\nonumber\\
&+& \frac{2[1-Dr^2 \pm \sqrt{(1-Dr^2)^2 - Kr^4}]}
{\beta^2[(1-Dr^2)^2 -Kr^4]}\,dr^2 + r^2(d\theta^2 + \sin^2 \theta
\,d\phi^2)\,,
\label{type1}
\end{eqnarray}
\begin{eqnarray}
ds^2= -\frac{dt^2}{G +\sqrt{r^{-2}+K}} + \frac{dr^2}{\beta^2(1+Kr^2)}
+r^2(d\theta^2 + \sin^2 \theta\,d\phi^2)\,,  
\label{type2}
\end{eqnarray}
which will be called the Type I and Type II forms of the BST respectively. The parameters $D$, $G$ and $K$ are
real, and $\beta$ must be a positive rational number.
  
For mathematical simplicity, we will consider BSTs of type II,
defined by the metric of eq.(\ref{type2}).  We wish to first
understand what type of matter distribution can cause this metric,
assuming the Einstein equations to hold. To this end, we construct the
Ricci scalar and the energy momentum tensor. The general expressions
are too lengthy to reproduce here, and we will frame our arguments
based on special cases. Let us begin with the case $K=0$ for which the
type II metric of eq.(\ref{type2}) reduces to
\begin{equation}
ds^2 = -\frac{dt^2}{G +r^{-1}} + \frac{dr^2}{\beta^2}
+r^2(d\theta^2 + \sin^2 \theta\,d\phi^2)\,,  
\label{type2a}
\end{equation}
where we will take $G >0$ to ensure a Lorentzian metric. The Ricci scalar can be calculated to
be \footnote{Here and in the rest of this section, we set $\theta =
  \frac{\pi}{2}$ in the final expressions, without loss of generality.}
\begin{equation}
R = \frac{4\left(1 + Gr\right)^2- \beta^2\left(4Gr\left(2 + Gr\right) +7\right)}{2r^2\left(1 + Gr\right)^2}
\end{equation}
this diverges at $r \to 0$ and vanishes as $r \to \infty$. The stress-energy tensor is proportional to 
\begin{equation}
T^{\mu\nu} = {\rm diag}\left(
\frac{\left(1-\beta^2\right) (G r+1)}{r^3}, \frac{\beta^2 \left((G r+2) 
\beta^2-G r-1\right)}{r^2 (G r+1)}, \frac{\beta^2 (1-2 G r)}
{4 r^4 (G r+1)^2},\frac{\beta^2 (1-2 G r)}{4 r^4 (G r+1)^2}\right)
\label{tmunuk0}
\end{equation}
To analyze the energy conditions, it is convenient to introduce an orthonormal frame that
satisfies
\begin{equation}
g_{\mu\nu}e^{\mu}_{\alpha}e^{\nu}_{\beta} = \eta_{\alpha\beta}
\end{equation}
where $\eta_{\alpha\beta} = {\rm diag}\left(-1,1,1,1\right)$ is the
flat Lorentzian metric. Since the metric of eq.(\ref{type2a}) is
diagonal, it is easy to see that a choice of the orthonormal basis is
given by $e^{\mu}_{\alpha} = {\rm diag}\left(\frac{1}{{\sqrt -
    g_{00}}},\frac{1}{{\sqrt g_{11}}},\frac{1}{{\sqrt
    g_{22}}},\frac{1}{{\sqrt g_{33}}}\right)$ whence the energy
momentum tensor can be written as
\begin{equation}
T^{\mu \nu} = \rho e^{\mu}_0e^{\nu}_0 + p_1e^{\mu}_1e^{\nu}_1 + p_2e^{\mu}_2e^{\nu}_2 + p_3e^{\mu}_3e^{\nu}_3
\label{tmunugen}
\end{equation}
and the energy density $\rho$ and the principal pressures $p_i,~i=1\cdots 3$ are 
\begin{equation}
\rho = \frac{1-\beta^2}{r^2},~p_1 = \frac{\beta^2\left(2+Gr\right) - \left(1+Gr\right)}{r^2\left(1+Gr\right)},~~
p_2 = p_3 = \frac{\beta^2\left(1-2Gr\right)}{4r^2\left(1+Gr\right)^2}
\end{equation}

It is seen that $\beta > 1$ is ruled out on physical grounds. $\beta = 1$ is somewhat unphysical,
as it implies a vanishing energy density in the presence of non zero pressures. For
$\beta \to 1^-$, the weak energy condition \cite{wald}, \cite{poisson}, 
$\rho \geq 0$, $\rho + p_i \geq 0$, $i=1,\cdots,3$ is
satisfied for $r < \frac{1}{2G}$. The WEC provides an interesting
upper bound on $r$, and $G$ has to be a small positive number for a
physically meaningful solution in this case. For $\beta < 1$, the WEC
is satisfied for certain intervals of $r$, depending on the choice of
$G$. Specifically, it can be checked that for positive values of G
(necessary to retain the Lorentzian nature of the metric of
eq.(\ref{type2a}) at large values of $r$), the WEC is satisfied for
all $r$.

Let us now turn our attention to non-zero values of $K$, where the situation is more complicated. With $K\neq 0$, the Ricci scalar diverges at 
$r \to 0$, and in the limit $r \to \infty$, $R_{\infty} = -6K\beta^2$.
The energy density and the principal pressures can be found by introducing an orthonormal frame analogous to the case $K=0$, and we find that 
\begin{equation}
\rho = \frac{1-\beta^2 \left(3 K r^2+1\right)}{r^2}
\label{rhokneq0}
\end{equation}
For $r \gg 1$, this implies that the energy density is negative for positive values of $K$. The situation might be remedied by choosing a negative value of $K$,
but note that this necessitates, from eq.(\ref{type2}) that for $K = -\kappa$ where $\kappa$ is a positive real number, we require $r < 1/\sqrt{\kappa}$. We can thus
choose $\kappa \ll 1$ so that the positivity of the energy density of space-time of eq.(\ref{type2}) is guaranteed for a large range of $r$. The analysis of the WEC
is similar to the case $K=0$ considered earlier. We will omit the algebraic details here and simply state the result that setting $\beta = 1$ for simplicity, 
for a given choice of $\kappa$ (in accordance with the discussion above), the WEC is always satisfied for $r < 1/\sqrt{\kappa}$.

Before we end this section, we will briefly comment on the strong energy condition \cite{wald},\cite{poisson} that follows from eq.(\ref{tmunugen}) :
$\rho + \sum_i p_i \geq 0,~~\rho + p_i \geq 0$. We find that for the metric of eq.(\ref{type2}),
\begin{equation}
\rho + \sum_ip_i = \frac{3 \beta^2}{2 r^2 \left(G r+\sqrt{K r^2+1}\right)^2}
\label{secbert}
\end{equation}
so that the SEC is satisfied  whenever $r < 1/\sqrt{\kappa}$, for positive values of $G$. This will be important for us in the next section. 

To summarize, in this section we have studied the energy conditions of Bertrand space-times of type II, given by the metric of 
eq.(\ref{type2}). An entirely similar analysis can be carried out for the Type I metric of eq.(\ref{type1}), although the algebraic expressions are 
complicated. We now proceed to study geodesic flows in BSTs.
 
\section{Geodesic Flows in Bertrand Space-times}
The kinematics of geodesic congruence in any space-time can be
specified by three quantities: the isotropic expansion, the
shear, and the rotation variables. In totality these are generally called the ESR variables, and the
evolution of these are guided by the Raychaudhuri equations
\cite{senguptakar}. Treating the geodesic congruence as a deformable fluid, one
can write the evolution equation of the vector between two fluid
points. This vector may get deformed as the geodesics flow, and
consequently the vector is called the deformation vector.  The
Raychaudhuri equations connect the evolution of the deformation vector
with the curvature of space-time. In general the deformation vector is
called $\xi^\mu$ and its rate of change with respect to an affine
parameter is given as
\begin{eqnarray}
\dot{\xi^\mu}=B^\mu_{\,\,\,\nu}\xi^\nu\,,
\label{xidot}
\end{eqnarray}
where the affine parameter interval in which the rate is measured is
supposed to be small. Here $B^\mu_{\,\,\,\nu}$ is a second rank tensor
characterizing the time evolution of the deformation vector,
\begin{eqnarray}
B^\mu_{\,\,\,\nu}=\nabla_\nu u^\mu\,,
\label{bmunu}
\end{eqnarray}
where $u^\mu$ is a tangent vector field which serves as the first
integral of the geodesic equations. Specifically, $u^\nu \nabla_\nu
u^\mu=0$, and choosing a suitable affine parameter one can make $u_\mu
u^\mu=-1$ for time-like geodesics while $u_\mu u^\mu=0$ for a null geodesic.
Differentiating the expression in Eq.~(\ref{xidot}) with respect to
the affine parameter one gets 
\begin{eqnarray}
\ddot{\xi^\mu}=(\dot{B}^\mu_{\,\,\,\nu} + B^\mu_{\,\,\,\tau}
B^\tau_{\,\,\,\nu})\xi^\nu\,.
\label{xiddot}
\end{eqnarray}
The Raychaudhuri equations are obtained by writing 
$\ddot{\xi^\mu}=-R^\mu_{\,\,\,\kappa \tau \nu}u^\kappa u^\nu\xi^\tau$ and
equating this with eq(\ref{xiddot}).

In $n$ space-time dimensions, the general form of the second rank tensor $B_{\mu
\nu}$ can be decomposed into irreducible parts as \cite{poisson}
\begin{eqnarray}
B_{\mu \nu}=\frac{1}{n-1}\Theta h_{\mu\nu} + \sigma_{\mu\nu} + \omega_{\mu\nu}\,,
\label{bdecomp}
\end{eqnarray}
where $h_{\mu\nu} = g_{\mu\nu} + u_{\mu}u_{\nu}$ for $u_{\mu}$ time-like,
and $\Theta$ is the expansion variable, $\sigma_{\mu\nu}$ is associated
with shear and $\omega_{\mu\nu}$ signifies rotation. The physical significance of these
variables are nicely explained in \cite{poisson}. One can explicitly write
\begin{eqnarray}
\Theta &=& B^\mu_{\,\,\,\mu}\,,
\label{thetaexp}\\
\sigma_{\mu\nu} &=& \frac12 (B_{\mu \nu} + B_{\nu \mu}) - \frac{1}{n-1}
\Theta h_{\mu\nu}\,,
\label{sigmaexp}\\
\omega_{\mu\nu} &=& \frac12 (B_{\mu \nu} - B_{\nu \mu})\,. 
\end{eqnarray}
and the ESR variables are generally denoted by $\Theta$, $\sigma^2$ and $\omega^2$. From
Eq.~(\ref{bmunu}) it is seen that if one knows the form of $u^\mu$ one
can calculate $B_{\mu \nu}$, and hence the ESR parameters.These are expected to 
give us information about geodesic flows and their properties in BSTs. 

\subsection{Geodesics in BSTs of Type II : General Considerations}
\label{geodbst}
We now focus on Type II BSTs, and further 
simplify the situation by choosing $\theta = \pi/2$, so that we are on the equatorial plane. In that
case we have the metric
\begin{eqnarray}
ds^2= -\frac{dt^2}{G +\sqrt{K + r^{-2}}} + \frac{dr^2}{\beta^2(1+Kr^2)}
+r^2d\phi^2\,.  
\label{stype2}
\end{eqnarray}
The geodesic equations can now be written down. The first one is
obvious from the form of the above line element,
\begin{eqnarray}
\frac{\dot{t}}{G + \sqrt{K + r^{-2}}}=C\,,
\label{tfsic}
\end{eqnarray}
where $C$ is a constant of integration. This equation can also be written as
\begin{eqnarray}
\ddot{t}+\frac{\dot{t}\dot{r}}{r^2\sqrt{1+Kr^2}(G+\sqrt{K+r^{-2}})}=0\,.
\label{tsic}
\end{eqnarray}
The other geodesic equations are:
\begin{eqnarray}
\ddot{r}+\frac{\beta^2\sqrt{1+Kr^2}\,\dot{t}^2}{2r^2(G+\sqrt{K+r^{-2}})^2}
-\frac{Kr \dot{r}^2}{1+Kr^2}-r\beta^2(1+Kr^2)\dot{\phi}^2=0\,,
\label{rsic} 
\end{eqnarray}
and
\begin{eqnarray}
\ddot{\phi}+\frac{2}{r}\dot{r}\dot{\phi}=0\,.
\label{psic}
\end{eqnarray}
On a radial geodesic Eq.~(\ref{tfsic}) holds and more over
$\dot{\phi}=0$. For a time-like geodesic ($u^\mu u_\mu=-1$), one can
write 
\begin{eqnarray}
-\frac{\dot{t}^2}{G+\sqrt{K+r^{-2}}}+\frac{\dot{r}^2}{\beta^2(1+Kr^2)}
+r^2\dot{\phi}^2=-1\,.
\label{usqr}
\end{eqnarray}
Using the above equation one can calculate the value of
$u^r=\dot{r}=dr/d\lambda$ on a radial geodesic. Here $\lambda$ is an
affine parameter. The value of $u^r$ comes out as
\begin{eqnarray}
u^r=\frac{dr}{d\lambda}=
\beta\sqrt{(1+Kr^2)\left[C^2\left(\sqrt{K+r^{-2}}+G\right)-1\right]}
\label{ur}
\end{eqnarray}
The above equation is for the outgoing radial geodesic directed away
from the origin. Note that, assuming $Kr^2 + 1 >0$ (see discussion in
section 2), this implies that there is a turning point of the outgoing
radial time-like geodesics, for $C^2\left(\sqrt{K + r^{-2}} + G\right)
= 1$. This implies that there is a maximum value of $r$ at which
outgoing radial geodesics stop.  This is analogous to the case of the
Schwarzschild black hole, where it is known that such turning points
occur for non-marginally bound radial time-like geodesics. The other
components of the tangent vector $u^\mu$ on the radial geodesic are
\begin{eqnarray}
u^t=\frac{dt}{d\lambda}=C\left(G + \sqrt{K+r^{-2}}\right)\,,
\,\,\,\,\,\,\,\,
u^\phi=\frac{d\phi}{d\lambda}=0\,.
\label{utup}
\end{eqnarray}
Having calculated the relevant components of the tangent vectors $u^\mu$ on
a radial time-like geodesic of Type II BSTs, one can compute the components of the $B^\mu_{\,\,\,\nu}$ tensor for radial time-like geodesics. 

For future reference, let us also list the components of the tangent vector for the radial null and the circular time-like geodesics. For the former, we
obtain
\begin{eqnarray}
u^t &=& C \left(G +\sqrt{K +r^{-2}}\right)\,,\nonumber\\
u^r &=& \beta C\sqrt{\left(Kr^2 + 1\right)\left(G + \sqrt{K + r^{-2}}\right)}\nonumber\\
u^\phi &=&0\,,
\label{radialnull}
\end{eqnarray}
where $C$ is defined in Eq.~(\ref{tfsic}). It is interesting to note
that for null radial geodesic case there is no turning point (for
$G>0$ as is always assumed in this article) as was present for the
time-like radial geodesics. This implies that light propagating away
along the radial direction in BST of type II is not bound to return
after travelling a finite distance. In the later part of this article
we will see that although outgoing null radial geodesics do not have a turning
point, an outgoing radial null geodesic congruence does focus away from the origin.

For the circular time-like geodesics, we find
\begin{eqnarray}
u^t &=&\sqrt{2} r \sqrt{\frac{\sqrt{K+r^{-2}} \left(G+\sqrt{K+r^{-2}}\right)^2}{2 r \left(G \sqrt{K r^2+1}+K r\right)+1}}\nonumber\\
u^r&=&0\,,\nonumber\\
u^\phi&=&\sqrt{\frac{1}{r^2 \left[2 r \left(G \sqrt{K r^2+1}+K r\right)+1\right]}}
\end{eqnarray}
From the above expressions of the tangent vectors one can see that
there exists an upper bound on the radial distance up to which BST of
type II can accommodate time-like circular geodesics. The upper limit
is given by the inequality
\begin{eqnarray}
2 r \left(G \sqrt{K r^2+1}+K r\right)+1 > 0\,.
\label{ineq}
\end{eqnarray}
If $K=0$ the above inequality is satisfied for all $r$. On the other
hand if $G=0$ and $K=-\kappa$ where $\kappa >0$, the upper bound is
given by $1/\sqrt{2\kappa}$. In general when  $G>0$ and $K<0$ the
upper bound on $r$ has to be evaluated by solving the
inequality in eq.(\ref{ineq}). If both $G$ and $K$ are greater than
zero the inequality in eq.(\ref{ineq}) is trivially satisfied but as
we have seen in section \ref{bst} that in this case the WEC and SEC
are violated for $r \gg 1$. 

To summarize, in this subsection, we have considered the BST of type
II (eq.(\ref{type2})), and calculated the first integrals of the
geodesic equation for radial time-like, radial null and circular
time-like vectors. These can be used in a straightforward manner to
evaluate the ESR parameters for BSTs, which we now turn to.

\subsection{The ESR variables for Type II BSTs}
\label{bmunucomp}
In this subsection, we compute the ESR parameter for BSTs of type II. Consider first the radial time-like geodesics. We start from eq.(\ref{bmunu}), from which we can write
\begin{eqnarray}
B^\mu_{\,\,\,\nu}=\frac{\partial u^\mu}{\partial x^\nu}
+\Gamma^\mu_{\nu \rho}u^\rho\,,
\label{nbmunu}
\end{eqnarray}
The non-zero components of $B^\mu_{\,\,\,\nu}$, required to evaluate the ESR variables
for the radial time-like geodesic flow, in the equatorial plane 
for the Type II Bertrand space-time, are listed below:
\begin{eqnarray}
B^t_{\,\,\,t}&=&\frac{\beta\sqrt{C^2(G+\sqrt{K+r^{-2}})-1}}
{2r^2(G+\sqrt{K+r^{-2}})},~~
B^t_{\,\,\,r}=\frac{C}{2r^2\sqrt{1+Kr^2}}\nonumber\\
B^r_{\,\,\,t}&=&-\frac{\beta^2 C \sqrt{1+Kr^2}}
{2r^2(G+\sqrt{K+r^{-2}})},~~
B^r_{\,\,\,r}=-\frac{\beta
C^2}{2r^2\sqrt{C^2(G+\sqrt{K+r^{-2}})-1}}\nonumber\\
B^\phi_{\,\,\,\phi}&=&\frac{\beta}{r}\sqrt{(1+Kr^2)[C^2(G+\sqrt{K+r^{-2}})-1]}\,.
\label{bmunu3}
\end{eqnarray}
Now from eq.(\ref{thetaexp}), we get 
\begin{eqnarray}
\Theta=\frac{\beta\sqrt{1+Kr^2}(Gr+\sqrt{1+Kr^2})
\left[2C^2 - \frac{r(3+2Kr^2+2Gr\sqrt{1+Kr^2})}{\sqrt{1+Kr^2}(Gr+
\sqrt{1+Kr^2})^2}\right]}{2r^2\sqrt{C^2(G+\sqrt{K+r^{-2}})-1}}\,.
\label{theta}
\end{eqnarray}
The shear coefficient squared, $\sigma^2 \equiv \sigma_{\mu\nu}\sigma^{\mu\nu}$,
for the radial time-like geodesics comes out as
\begin{eqnarray}
\sigma^2=\frac{\beta {\mathcal P}(r)}{{\mathcal Q}(r)}\,,
\label{sig}
\end{eqnarray}
where ${\mathcal P}(r)$ and ${\mathcal Q}(r)$ are are functions of $r$, given by 
\begin{eqnarray}
{\mathcal P}(r) &=&
2 K r^3 \left(2 C^2 G-1\right)+2 r^2 \sqrt{K r^2+1} \left(G \left(C^2 G-1\right)+C^2 K\right)\nonumber\\
&~& +r \left(4 C^2 G-3\right)+2 C^2 \sqrt{K r^2+1}\nonumber\\
{\mathcal Q}(r) &=&
2 r^2 \left(G r+\sqrt{K r^2+1}\right) \sqrt{C^2 \left(G+\frac{\sqrt{K r^2+1}}{r}\right)-1}
\label{pr}
\end{eqnarray}
The rotation parameter for the radial time-like geodesics,
$\omega^2\equiv \omega_{\mu\nu}\omega^{\mu\nu}=0$.  Before we move on, let us make a few comments. First of
all, note that $\Theta$ and $\sigma^2$ diverge at $r=0$, corresponding
to the singularity of the BST at that point.  These also diverges at
$C^2\left(\sqrt{K + r^{-2}} + G\right) = 1$, the turning point for
outgoing radial time-like geodesics (see discussion after
eq.(\ref{ur})) and indicates that the geodesics focus or de-focus at
the turning point.

Next we present the ESR parameters for the circular time-like
geodesics. For these, the $B^\mu_{\,\,\,\nu}$ components are given as
\begin{eqnarray}
B^t_{\,\,\,r} &=& -\frac{(G+2GKr^2+2Kr\sqrt{1+Kr^2})}
{\sqrt{2r}(1+Kr^2)^{3/4}(1 + 2Kr^2 + 2Gr\sqrt{1+Kr^2})^{3/2}}
\label{ptr}\nonumber\\
B^r_{\,\,\,t} &=& \frac{\beta^2(1+Kr^2)^{3/4}}{\sqrt{2r}
(Gr+\sqrt{1+Kr^2})(1+2Kr^2+2Gr\sqrt{1+Kr^2})^{1/2}}\nonumber\\
B^r_{\,\,\,\phi} &=& -\frac{\beta^2(1+Kr^2)^{3/4}}{(1+2Kr^2+
2Gr\sqrt{1+Kr^2})^{1/2}}\,,
\label{prp}\nonumber\\
B^\phi_{\,\,\,r} &=& -\frac{(G+2GKr^2+2Kr\sqrt{1+Kr^2})}
{r\sqrt{1+Kr^2}(1 + 2Kr^2 + 2Gr\sqrt{1+Kr^2})^{3/2}}\,.
\label{ppr}
\end{eqnarray}
and the diagonal elements of $B^\mu_{\,\,\,\nu}$ are all
zero. This implies the expansion coefficient $\Theta=0$ for the
circular time-like geodesics. The shear coefficient is
\begin{eqnarray}
\sigma^2 = \frac{\beta^2 [\sqrt{1+Kr^2}(1+4Kr^2)+Gr(3+4Kr^2)]^2}
{4r^2\sqrt{1+Kr^2}(Gr+\sqrt{1+Kr^2})(1 + 2Kr^2 + 2Gr\sqrt{1+Kr^2})^{2}}\,.
\label{pss}
\end{eqnarray}
The rotation parameter $\omega^2$ for the circular time-like geodesics is
\begin{eqnarray}
\omega^2=\frac{\beta^2(Gr+\sqrt{1+Kr^2})}{4r^2\sqrt{1+Kr^2}
(1 + 2Kr^2 + 2Gr\sqrt{1+Kr^2})^{2}}\,.
\label{poms}
\end{eqnarray}
To discuss the behavior of these parameters, let us first consider the
case $K=0$. In this case, the shear and rotation parameters reduce to 
\begin{eqnarray}
\sigma^2 &=& \frac{\beta ^2 (3 G r+1)^2}{4 r^2 (G r+1) (2 G r+1)^2}\nonumber\\
\omega^2 &=&  \frac{\beta ^2 (G r+1)}{4 r^2 (2 G r+1)^2}
\end{eqnarray}
These are positive everywhere, for any value of $\beta$ and go to zero
in the limit of infinite $r$ (remember that $G$ is positive). For
$G>0$ and $K<0$ the analysis of the shear and the rotation
parameter cannot be simply guessed by the expressions in
eqs.(\ref{pss}) and (\ref{poms}) as in this case one can have an upper
bound of the radial distance up to which circular time-like geodesics
can be obtained in Type II BST.

Finally, let us briefly discuss null geodesics. 
For the radial null geodesics the evolution tensor $\tilde{B}_{\mu \nu}$ is 
defined as
\begin{eqnarray}
\tilde{B}_{\mu \nu}=P^\lambda_{\,\,\,\,\mu}B_{\lambda \kappa}
P^\kappa_{\,\,\,\,\nu}\,,
\label{btilde}
\end{eqnarray}
where the projection tensor $P_{\alpha \beta}$ is 
\begin{eqnarray}
P_{\alpha \beta}=g_{\alpha \beta} + n_\alpha u_\beta + n_\beta u_\alpha\,.
\label{project}
\end{eqnarray}
Here $u_\alpha$ is the first integral of null geodesic equations
i.e., $u^\beta \nabla_\beta u_\alpha=0$ and $u_\alpha u^\alpha=0$.
The vector $n_\alpha$ satisfies the following conditions:
\begin{eqnarray}
n_\alpha n^\alpha = 0\,,\,\,\,\,n_\alpha u^\alpha = -1\,,\,\,\,\,
u^\beta \nabla_\beta n_\alpha=0\,.
\label{conds}
\end{eqnarray}
Now the evolution tensor (in an effectively one dimensional space) becomes
\begin{eqnarray}
\tilde{B}_{\mu \nu}=\Theta P_{\mu \nu} + \tilde{\sigma}_{\mu \nu}
+ \tilde{\omega}_{\mu \nu}\,,
\label{dbtilde}
\end{eqnarray}
where
\begin{eqnarray}
\Theta &=& \tilde{B}^\mu_{\,\,\,\,\mu} = B^\mu_{\,\,\,\,\mu}\,,\nonumber\\ 
\tilde{\sigma}_{\mu \nu} &=& \frac12(\tilde{B}_{\mu \nu}+\tilde{B}_{\nu \mu})
- \Theta P_{\mu \nu}\,,\nonumber\\
\tilde{\omega}_{\mu \nu} &=& \frac12 (\tilde{B}_{\mu \nu} -
\tilde{B}_{\nu \mu})\,.
\end{eqnarray}
The vector $n^\alpha$ is tangent to
the inward radial geodesics. If we take
\begin{eqnarray}
n^t &=& -\frac{1}{2 C}\,,\nonumber\\
n^r &=& -\frac{\sqrt{r} \beta C\sqrt{\left(K r^2+1\right) 
\left(G r+\sqrt{K r^2+1}\right)}}
{2 C^2 \left(G r+\sqrt{K r^2+1}\right)}\,,\nonumber\\
n^\phi &=& 0\,,
\end{eqnarray}
then these components  $n^\alpha$ satisfy all the conditions in 
Eq.~(\ref{conds}). Using these we can find all the components of 
$P^\lambda_{\,\,\,\,\mu}$. The only non zero component of
$P^\lambda_{\,\,\,\,\mu}$ turns out to be $P^\phi_{\,\,\,\,\phi}$
which is given as
\begin{eqnarray}
P^\phi_{\,\,\,\,\phi} = 1\,.
\end{eqnarray}
Using the above component of the projection tensor one finds that there is only one non-zero element of $\tilde{B}^\mu_{\,\,\,\,\mu}$, namely, $\tilde{B}^\phi_{\,\,\,\,\phi}$. 
Consequently, for the radial null geodesics, we obtain
\begin{equation}
\Theta = \tilde{B}^\phi_{\,\,\,\,\phi} = \beta Cr^{-\frac{3}{2}}\sqrt{\left(K r^2+1\right) 
\left(G r+\sqrt{K r^2+1}\right)}
\label{thetanull}
\end{equation}
The above expression of the expansion variable for the null radial
geodesic congruence for Type II BST shows that $\Theta$ has a
singularity at $r=0$ where BST of Type II itself is singular. Unlike
the outgoing time-like radial geodesics the expansion variable for the
outgoing null radial geodesics do not have any other singularity. 

\subsection{Analysis of BST Type II spacetime in terms of the ESR parameters}
Having obtained the ESR parameters for BSTs of type II, let us analyze
them in some detail.  First, we focus on radial time-like
geodesics. In this case, the ESR parameters can be obtained by setting
$K=0$ in eq.(\ref{theta}) and in eqs.(\ref{sig}), (\ref{pr}). As mentioned in the discussion after eq.(\ref{tmunuk0}),
we can consider two cases here, namely $\beta = 1$ and $\beta <
1$. Let us focus on the former case. Here, the expression for
the expansion parameters assumes a simple form
\begin{equation}
\Theta = \frac{(G r+1) \left(2 C^2-\frac{r (2 G r+3)}{(G
    r+1)^2}\right)}{2 r^2 \sqrt{C^2 \left(G+\frac{1}{r}\right)-1}}
\label{thetak0}
\end{equation}
with $\sigma^2$ given from eqs.(\ref{sig}) and (\ref{pr}), with $K=0$. 

\begin{figure}[t!]
\begin{minipage}[b]{0.5\linewidth}
\centering
\includegraphics[width=2.8in,height=2.3in]{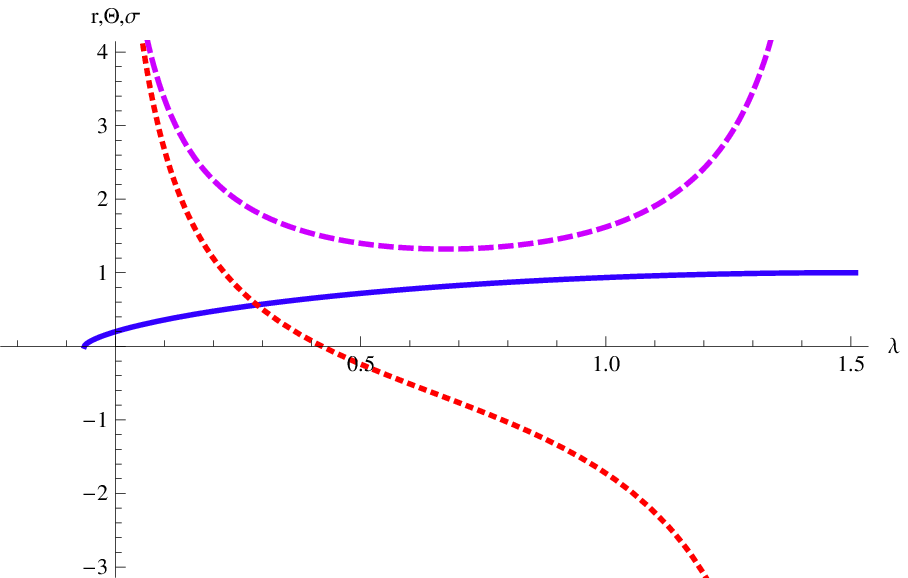}
\caption{Numerical solutions for $r$ (solid blue), $\Theta$ (dotted
  red), and $\sigma\equiv \sqrt{\sigma_{\mu \nu}\sigma^{\mu \nu}}$ 
(dashed magenta) for radial time-like geodesics in 
type II BST, as a function of $\lambda$ for $K=0$. For details, see text.}
\label{esrk0}
\end{minipage}
\hspace{0.2cm}
\begin{minipage}[b]{0.5\linewidth}
\centering
\includegraphics[width=2.8in,height=2.3in]{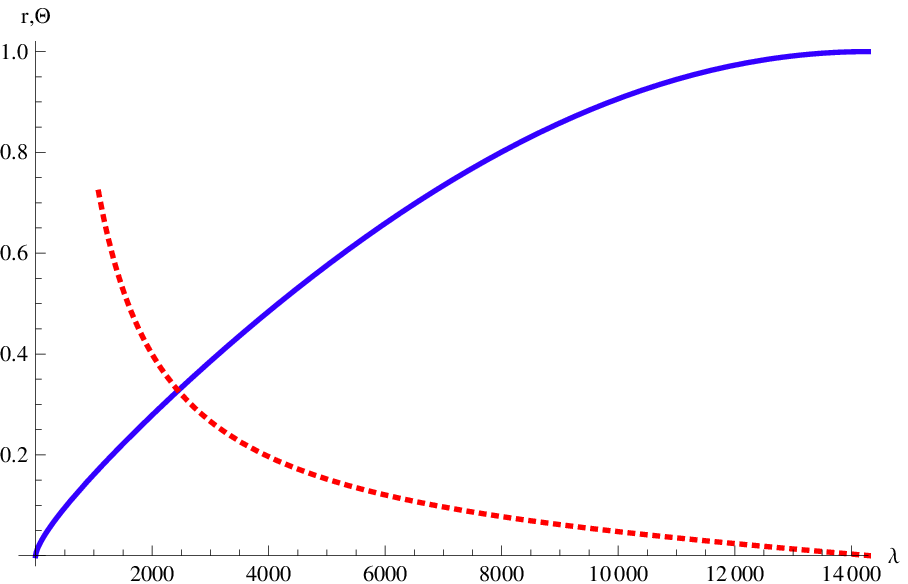}
\caption{Numerical solutions for $r$ (solid blue) and $\Theta$ (dotted red) for radial null geodesics in type II BST, as a function of $\lambda$ for the value
$K=-10^{-6}$. For details, see text.}
\label{esrkneq0}
\end{minipage}
\end{figure}

We wish to understand the behavior of $\Theta$ and $\sigma$ as a
function of the affine parameter, $\lambda$. (Note, however that in
our method, no initial condition on the parameter $\Theta$ or $\sigma$
is possible, unlike the case where one integrates the full set of
Raychaudhuri equations with given initial conditions
\cite{sayan1},\cite{sayan2} (see also \cite{sayan3}). We will proceed, keeping this in
mind). To this end, we first numerically solve eq.(\ref{ur}) (with
chosen upper and lower limits of the affine parameter) to express the
radial coordinate $r$ as a function of $\lambda$. This is then fed
back in eq.(\ref{thetak0}) to obtain the behavior of $\theta$ as a
function of $\lambda$. For illustration purpose, we have chosen
$G=10^{-3}$, and set $C=1$, and the lower and upper limits of the
affine parameter $\lambda$ have been set to $-0.2$ and $1.5$. The
upper limit of $\lambda$ is chosen so that $r$ varies from zero to the
turning point of $u^r$, which can be seen to be $r \sim 1$ in this
case (these numbers are simply for illustration).  The result is shown
in fig.(\ref{esrk0}), where the solid blue, dotted red and the dashed
magenta curves correspond to numerical solutions for $r$, $\Theta$ and
$\sigma$ respectively, as a function of the affine parameter, with the
chosen initial condition. From the figure, we see
that $\frac{d\theta}{d\lambda}$ is always negative, confirming the
focusing theorem \cite{wald},\cite{poisson} given the validity of the SEC of eq.(\ref{secbert}). Also, $\Theta$ diverges at the upper and lower
limits of $r$, signaling the turning points of the geodesics. We find
that the case $K \neq 0$ for radial time-like geodesics follow the
same qualitative behavior.

To illustrate the case of null geodesics, we have taken
$K\neq 0$. We have followed a numerical procedure similar to that
alluded to above, and chosen $K = -10^{-6}$, $G = 10^{-2}$, and $C$
and $\beta$ has been set to unity. Here, the lower and upper limits of
the affine parameter has been set to $-0.2$ and $1.389$
respectively. Using the same numerical procedure as above, we solve
for the expansion parameter of eq.(\ref{thetanull}), and this is
illustrated in fig.(\ref{esrkneq0}), where we have multiplied $\Theta$
by a factor of $10^3$ to display the curves on the same
graph. Note that in this case the expansion parameter diverges at
$r=0$ as expected from eq.(\ref{thetanull}).

Before we conclude, we briefly mention the expansion parameter for the
BST of type II in the four dimensional case. The details are
unimportant here, and these can be worked out exactly like the case $\theta = \frac{\pi}{2}$. 
For radial time-like geodesics, the
expansion parameter is $\Theta = \beta{\mathcal
  A}(r)/{\mathcal B}(r)$ where
\begin{eqnarray}
{\mathcal A}(r) &=& (4 r^2 \sqrt{K r^2+1} \left(C^2 G^2+C^2 K-G\right)+4 K r^3 \left(2 C^2 G-1\right)\nonumber\\
&+& r \left(8 C^2 G-5\right)+4 C^2 \sqrt{K r^2+1}\nonumber\\
{\mathcal B}(r) &=& 2 r^2 \left(G r+\sqrt{K
 r^2+1}\right) \sqrt{C^2 \left(G+\frac{\sqrt{K r^2+1}}{r}\right)-1}
\end{eqnarray}
For radial null geodesics in four dimensions, the expansion
parameter turns out to be twice that of eq.(\ref{thetanull}). The
shear parameter for the radial time-like case yields a lengthy
expression analogous to eqs.(\ref{sig}) and (\ref{pr}), which we omit for brevity, and
is zero for the radial null case, as before.
\section{Conclusions and Discussions}
In this paper, we have considered the energy conditions and geodesic
deformations in Bertrand space-times. For simplicity, we have chosen
the BST of type II (eq.(\ref{type2})), although we believe that the
qualitative results will remain unchanged even for a type I BST.  As
the metric of BST of type II contains three parameters we had to check
the probable ranges of these parameters which gives the spacetime 
physical significance. We have explicitly checked the weak and strong
energy conditions for type II BSTs, and verified their validity within
certain ranges of parameters. The ESR parameters for both time-like
and null geodesics in the BST type II spacetime are calculated as
functions of the radial coordinate. For simplicity most of the
calculations are done in the equatorial plane. We have not explicitly
solved the Raychaudhuri equation in the above mentioned spacetime but
indirectly obtained its solution by expressing the radial coordinate
in terms of the affine parameter in the ESR variables. In such a
situation the ESR variables becomes functions of the affine parameter
and these variables are now valid solutions of the Raychaudhuri
equation for geodesic deformations in type II BSTs.  While
analyzing the solutions of the Raychaudhuri equation for geodesic
deformations in type II BSTs we have confirmed the focusing theorem
numerically.

The behavior of the solution of the Raychaudhuri equation for
geodesic deformations in type II BST for the time-like geodesic
congruence is analyzed in this article with some suitable choice of
parameter values $K$, $G$ and $\beta$. In the specific case chosen it
is seen that generally the outgoing radial time-like/null geodesics
diverge from $r=0$, which is a singular point in this spacetime where
the Ricci scalar diverges. The outgoing radial time-like geodesics do
not diverge indefinitely, they do converge again at some other value
of $r$ which turns out to be a turning point in type II BST. Where as
an outgoing radial null geodesic congruence do not converge at any
finite $r$ but it does show focusing property.  At the turning point
for the time-like radial geodesics, spacetime is not singular but the
radial component of the tangent vector to the radial geodesics vanish
at this point. For the radial geodesics the rotation parameter
is always zero but the time-like radial geodesics do show extreme
shear near $r=0$ and the turning point. The circular time-like
geodesics do not show any focusing behavior, as expected. But the
circular orbits do show rotation.  

Our analysis points to the fact that BSTs can be thought of as
interesting realistic examples of static, spherically symmetric
space-times, which are asymptotically non-flat, and allow for stable,
periodic orbits. This might be significant in astrophysical scenarios
: for example, one might ask if a realistic space-time near a compact
object can be modeled via BST of type II or I. In this article the
main attention was given to the geodesics of BST of type II and its
ESR variables to understand the effect of spacetime curvature and
probable singularities. To properly utilize BSTs, one must also have
to think of the source of such kind of spacetime's and in future works
one needs to look at this important aspect.  For completeness, it will
be of interest to study, in the same manner, BSTs of type I and
analyze the parameter space of this theory. We leave this for a future
publication.
\begin{center}
{\bf Acknowledgements}
\end{center}
It is a pleasure to thank Sayan Kar for several useful correspondences. 


\end{document}